\documentclass[12pt]{article}

\usepackage{amsmath,amssymb,mathrsfs}
\usepackage{graphicx}
\usepackage{caption}
\usepackage{color}
\usepackage{dcolumn}
\usepackage{bm}
\usepackage{float}
\usepackage{cite}
\usepackage{listings}
\usepackage{hyperref} 
\usepackage{titling}
\hypersetup{ colorlinks = true, citecolor = black, urlcolor = blue}

\usepackage[capitalize]{cleveref}
\crefname{figure}{Fig.}{Figs.}
\Crefname{figure}{Figure}{Figures}
\crefname{table}{Tab.}{Tabs.}
\Crefname{table}{Table}{Tables}
\crefname{equation}{Eq.}{Eqs.}
\Crefname{equation}{Equation}{Equations}
\crefname{section}{Sec.}{Secs.}
\Crefname{section}{Section}{Sections}

\definecolor{background-color}{gray}{0.98}

\newcommand{\xcref}[0]{\cite{Liu05_241102,Peng06_044102,Saue07_064102,Reiher13_184105,Cheng07_104106,Liu16_204}}
\newcommand{\blasref}[0]{\cite{Krogh79_TOMS308,Hanson88_TOMS18,Duff90_TOMS18,Whaley02_TOMS135}}

\newcommand{\cqweb}[0]{\href{http://www.chronusquantum.org}{http://www.chronusquantum.org}}
\newcommand*\samethanks[1][\value{footnote}]{\footnotemark[#1]}

\thanksmarkseries{alph}

\title{The Chronus Quantum (ChronusQ) Software Package}

\author{
David B. Williams--Young
\thanks{Computational Research Division, Lawrence Berkeley National Laboratory, Berkeley, CA, 94720}
\thanks{Department of Chemistry, University of Washington, Seattle, WA, 98195, United States} 
\thanks{\color{red} Corresponding Author: dbwy@lbl.gov}~,
Alessio Petrone
\thanks{Dipartimento di Scienze Chimiche, Università di Napoli ‘Federico II’, Complesso Universitario di M.S. Angelo, via Cintia, I-80126 Napoli, Italy} , 
Shichao Sun \samethanks[2]~,\\ 
Torin F. Stetina  \samethanks[2]~, 
Patrick Lestrange \samethanks[2]~,
Chad E. Hoyer \samethanks[2]~, \\
Daniel R. Nascimento
\thanks{Department of Chemistry and Biochemistry, Florida State University, Tallahassee, FL 32306, United States} , 
Lauren Koulias \samethanks[2]~, 
Andrew Wildman \samethanks[2]~, \\
Joseph Kasper \samethanks[2]~,
Joshua J. Goings
\thanks{Department of Chemistry, Yale University, 225 Prospect Street, New Haven, CT, 06520, United States} , 
Feizhi Ding
\thanks{Division of Chemistry and Chemical Engineering, California Institute of Technology, Pasadena, CA, 91125, United States} , \\
A. Eugene DePrince III
\samethanks[5]~,
Edward F. Valeev
\thanks{Department of Chemistry, Virginia Tech, Blacksburg, VA 24061, United States} , 
Xiaosong Li \samethanks[2]
\thanks{\color{red} Corresponding Author: xsli@uw.edu}
}

\date{}
\begin{document}
\maketitle

\begin{center}
\subsubsection*{\small Article Type:}
Software Focus


\subsubsection*{Abstract}
\begin{flushleft}
The Chronus Quantum (ChronusQ) software package is an open source (under the
GNU General Public License v2) software infrastructure which targets the
solution of challenging problems that arise in \emph{ab initio} electronic
structure theory.  Special emphasis is placed on the consistent treatment of
time dependence and spin in the electronic wave function, as well as the
inclusion of relativistic effects in said treatments. In addition, ChronusQ
provides support for the inclusion of uniform finite magnetic fields as external perturbations
through the use of gauge-including atomic orbitals (GIAO).
ChronusQ is a parallel electronic structure code written in modern C++ which
utilizes both message passing (MPI) and shared memory (OpenMP) parallelism. 
In
addition to the examination of the current state of code base itself, a
discussion regarding ongoing developments and developer contributions will also
be provided.
\end{flushleft}
\end{center}

\clearpage

\renewcommand{\baselinestretch}{1.5}
\normalsize

\clearpage

\section*{\sffamily \Large INTRODUCTION} 

The development of the Chronus Quantum (ChronusQ) software package \cite{chronusq_beta2} began in the
research group of Xiaosong Li at the University of Washington in the beginning
of 2014. The primary purpose of ChronusQ at that time was to serve as an outlet
for the dissemination of some of the more ``unconventional" electronic
structure method development taking place in the group. Chief among these
developments was methodological experimentation in the fields of time-dependent
and relativistic electronic structure theory, and specifically at their
intersection.  Over the years, ChronusQ has received contributions from many
developers, and new and exciting functionality continues to be added to its
code base to this day.  ChronusQ is \emph{free} software released under the GNU
General Public License (v2), and is made publicly available at
\cqweb.  In this context, \emph{free} software is
meant to indicate that ChronusQ is \emph{gratis}, as well as the fact that it
grants its recipients the freedom to use, modify and redistribute the code
base under the same license.

The set of electronic structure methods implemented in the ChronusQ program
spans both the novel, and the tried and tested. Contemporary software which
implement similar  pieces of quantum chemistry functionality include the
ReSpect\cite{ReSpect}, DIRAC\cite{DIRAC18}, DALTON\cite{daltonpaper}, and
BAGEL\cite{BAGEL} programs. However there are a number of electronic structure
methods implemented in ChronusQ which are, at the time that this article has been
written and to the best of the authors' knowledge,
not currently available in any other public, open source quantum chemistry
codes. These methods will be highlighted in the main text.

As the name might imply (\emph{Chronus} being the latinization of 
\emph{Time} personified in Greek mythology), the core of the functionality in
ChronusQ centers around time-dependent electronic structure theory. However, it
is through the lens of time-dependent phenomena that ChronusQ
also provides a slew of related capabilities such the treatment of
electronically excited states and molecular properties. ChronusQ provides
functionality to solve time-dependent electronic structure problems both in the
time and frequency domains.  In conjunction with 
the time-dependent electronic structure methods, 
special emphasis is placed on the treatment of electronic spin
throughout the code base, specifically in the context of broken spin--symmetry
solutions. As such, all of the electronic structure methods implemented in
ChronusQ fully support the use of two-component spinor wave functions.  In
addition to these core functionalities, ChronusQ supports a number of
complementary features which are not typically found in general purpose quantum
chemistry software. Relativistic effects are supported throughout the code
through the use of the exact two--component (X2C) method \xcref. Magnetic fields may
also be added as an external perturbation through the use of gauge including atomic
orbitals (GIAO)
\cite{London37_397,Ditchfield72_5688,Jorgensen91_2595,Helgaker08_154114,Teale17_3636,Tellgren18_184112,Li19_348}.  
All electronic structure methods implemented in ChronusQ are
based on the linear combination of atomic orbitals (LCAO) paradigm to construct
the electronic wave function using contracted Gaussian-type orbital (CGTO)
basis set expansions. Both generally contracted and segmented CGTO basis sets are supported.
ChronusQ is a single-reference quantum chemistry code
which utilizes single Slater determinant descriptions of the ground state
electronic wave function based on either Hartree-Fock (HF) or Kohn-Sham density
functional theory (KS). Using the Slater determinant as a reference wave function,
ChronusQ also supports various post-SCF methods such as the particle hole \cite{Yeager84_33,Casida95_Book,HeadGordon05_4009} 
and particle particle 
\cite{Yang13_224105,Yang13_18A522,Yang13_174110} 
random phase approximations. Additionally, the development
of many body methods, such as coupled cluster theory \cite{Bartlett82_1910}, are currently underway.

Since its inception, the primary design goal of ChronusQ has been to enable
rapid development of novel research in electronic structure theory
without sacrificing performance on modern parallel computing architectures.
With the exception of a few external dependencies, ChronusQ is a standalone
software infrastructure completely written in modern C++. The current release
of ChronusQ relies on the C++14 standard, while a number of the development
streams utilize features outlined in C++17.  To this end, extensive use of
modern C++ paradigms (SFINAE, polymorphism, etc.) and template metaprogramming
are leveraged throughout the code base to enable a level of abstraction which
allows developers to write expressive and efficient code.  For an examination
of some of the more salient aspects of the C++ design of ChronusQ, we refer the
reader to the Appendix.  ChronusQ utilizes a hybrid parallelism scheme to
improve scalability on large computing clusters which leverages both message
passing and shared memory protocols to achieve inter- and intranode
concurrency, respectively. Calculations involving thousands of CPU
cores and thousands of basis functions are regularly performed with
ChronusQ on some of the world's largest supercomputers \cite{Li17_4950,Li18_169}. ChronusQ is compatible
with any MPI-2 compliant message passing implementation, such as MPICH, OpenMPI
and Intel(R) MPI. To allow for the use of ChronusQ on work
stations which may not possess an MPI implementation, message passing
parallelism is an optional feature which may be enabled or disabled by the
build system. Shared memory parallelism is achieved through a combination of
the explicit use of OpenMP threads and reliance on threaded external libraries
such as those which implement the BLAS \blasref~ and LAPACK \cite{LAPACK}
standards for numerical linear algebra.

After several years of development, ChronusQ has become a full-fledged,
standalone, \emph{ab initio}  quantum chemistry software package which has
proven itself useful both as a development platform to rapidly prototype
electronic structure methods, as well as a robust production code capable of
tackling complex science problems \cite{Li17_4950,Li19_348,Li19_3162,Li19_174114,Li19_accepted}.  In this article, we briefly review the
design, features and current developments of the ChronusQ software package.

\section*{\sffamily \Large DESIGN AND NOVEL STRUCTURAL FEATURES}

\subsection*{\sffamily \large Input Specification}

ChronusQ utilizes the INI format for the specification of its input.
An example ChronusQ input file specification for the time propagation
of a single water molecule at the cc-pVTZ / X2C-B3LYP \cite{Frisch94_11623,Dunning89_1007}
level of theory in the presence of an electric dipole field may
be found in \cref{fig:cq_input}. ChronusQ packages a representative
subset of commonly used CGTO basis sets, including the Sapporo \cite{Koga03_85,Matsuoka95_27}
basis sets commonly used in relativistic calculations, with its
source code. Arbitrary CGTO basis sets may be utilized by specifying the
full path for the desired basis set file in the Gaussian94 file format.
For a detailed examination of the anatomy and options supported by
the ChronusQ input file, we refer the reader to the ChronusQ user manual
which may be found at \cqweb.

\begin{figure}[hbtp]
\lstset{
                basicstyle=\tiny\ttfamily,
                keywordstyle=\color{blue}\ttfamily,
                stringstyle=\color{red}\ttfamily,
                commentstyle=\color{magenta}\ttfamily,
                morecomment=[l][\color{magenta}]{\#},
                frame=single
}
\begin{center}
\begin{tabular}{c}
\begin{lstlisting}[linewidth=6.9cm]
[Molecule] # Molecule Specification 
charge = 0
mult = 1
geom: 
  O     0.000000000  -0.0757918435  0.000000000
  H     0.866811829   0.6014357793  0.000000000
  H    -0.866811829   0.6014357793  0.000000000
[QM] # Job Specification
reference = X2CB3LYP
job = RT
[BASIS] # Basis Set Specification
basis = cc-pVTZ
[RT] # Time-Propagation Parameters
TMAX   = 1.0
DELTAT = 0.05
FIELD:
  StepField(0.00, 0.10) Electric 0.000 0.001 0.000
  StepField(0.05, 0.80) Electric 0.008 0.000 0.000
\end{lstlisting} 
\end{tabular}
\end{center}
\caption{An example specification for a RT ChronusQ job. The
field specified will add a $y$ electric dipole operator of
strength 0.001 AU on $t\in [0,0.1)$, and a $x$ electric dipole
operator of strength 0.008 AU on $t\in[0.05,0.8)$.}
\label{fig:cq_input}
\end{figure}

\subsection*{\sffamily \large Software Dependencies}

While ChronusQ is best described as a \emph{standalone} quantum chemistry
software suite, this description is primarily limited to the electronic
structure capabilities of the software. ChronusQ relies on a number of external
libraries to provide complementary functionality which facilitate efficient
implementations of quantum mechanical methods. Where possible and practical,
these software dependencies are packaged with the source code itself; however a
number of these dependencies must be installed separately and discovered by
the build system upon configuration. A list and description of these dependencies
is outlined in the ChronusQ user manual. 

To ensure portability on many different software configurations across different
architectures, ChronusQ utilizes a CMake build system. The CMake build system
attempts to determine the existence and linkage of the various dependencies
as well as provide specifications of various build options which
influence the performance and configuration of ChronusQ such as debug and optimization
flags, enabling MPI bindings, etc. A comprehensive guide to the CMake build
system may be found in the ChronusQ user manual.

ChronusQ relies extensively on optimized and threaded implementations of BLAS
and LAPACK numerical linear algebra standards to ensure high-levels of
performance on modern computing architectures. The open source OpenBLAS \cite{Yi13_ICHP1,Yunquan12_IEEE684}
library is packaged with ChronusQ to ensure that such performance is portable
to a large number of different architectures. ChronusQ is also compatible with
vendor tuned linear algebra software such as the Intel(R) Math Kernel Library
(MKL) and IBM Engineering and Scientific Software Library (ESSL) for specific
architectures.  If MPI has been enabled, ChronusQ may be optionally configured
to utilize the ScaLAPACK \cite{ScaLAPACK} library to perform high-performance distributed memory
linear algebra suitable for large computing clusters.

With the exception of those integrals which are specific to the X2C and GIAO
methods implemented in ChronusQ, molecular integrals over contracted Gaussian
basis functions are evaluated using the Libint library \cite{Libint2}. ChronusQ was one of the
early adopters of the modern C++ interface of the Libint library which has
enabled the development of expressive and reusable integral drivers. For
the evaluation of exchange-correlation (XC) functionals (and their derivatives) in the 
KS implementation, ChronusQ utilizes the Libxc
library \cite{Burnus12_2272,Marques18_1}. To ensure portability of standardization of the ChronusQ checkpoint
and binary files, we utilize the HDF5 library.

\subsection*{\sffamily \large General Infrastructure for the Unified Treatment of Electronic Spin}

Perhaps the most common motif in the design of ChronusQ is in the general
manner with which it treats electronic spin throughout the code base. It is in
this context that the development of ChronusQ has benefited from the era in
which it has been written: spin was considered a priority from the beginning.
Traditionally, quantum chemistry codes are generally written from a ``bottom-up"
approach as it relates to the breaking of spin symmetries: spin restricted methods
(RS: $S^2$ and $S_z$ eigenfunctions) are written before spin unrestricted
(US: $S_z$ eigenfunctions) methods which are written before generalized spin
methods (GS: broken spin-symmetry), if such GS treatments are
implemented at all. If GS is to be neglected in a particular implementation,
this ``bottom-up" approach is logical as the Roothaan-Hall equations separate
into distinct (but coupled) eigenvalue problems in the binary spin basis \cite{Szabo12_book},
\begin{equation}
\text{RS / US:} \qquad F^\sigma(P^\alpha, P^\beta) C^\sigma = S C^\sigma E^\sigma, \quad \sigma \in \{\alpha, \beta\}.
\label{eq:rh_1c}
\end{equation}
Here $\alpha$ and $\beta$ are spin indices, $S$ is the basis overlap matrix,
and $F^\sigma$, $C^\sigma$, $P^\sigma$, and $E^\sigma$ are the spin dependent Fock matrix,
molecular orbital matrix, density matrix, and diagonal matrix of canonical orbital
eigenenergies, respectively.  Round brackets, $(\cdot)$, indicate a functional dependence.
However, GS takes the form of a single eigenvalue problem of twice the dimension \cite{Paldus03_67},
\begin{equation}
\text{GS:} \qquad F^{2C}(P^{2C}) C^{2C} = S C^{2C} E^{2C}, \label{eq:rh_2c} 
\end{equation}
where
\begin{equation}
X^{2C} = \begin{bmatrix} X^{\alpha\alpha} & X^{\alpha\beta} \\ X^{\beta\alpha} & X^{\beta\beta} \end{bmatrix}, \qquad
X \in \{ F, P \}.
\end{equation}
Generalizing a code base from RS to US requires the manipulation of one
additional density/Fock matrix whereas generalization of US to GS requires
the manipulation of one fewer density/Fock matrix. GS almost seems
artificial in this treatment as it more closely resembles RS in the context of
the manipulation of individual matrices but resembles US in the treatment of
multiple spin degrees of freedom.  Such a structure requires a significant
amount of logic in the code base to control the proper workflow and
specialization of many aspects of the code to either work with RS/US or
GS.  In ChronusQ, the treatment of spin was a primary consideration from the
initial development, thus some time was spent in developing a unified code
infrastructure that treats RS, US, and GS on an equal footing. To
this end, we adopt a ``top-down" approach to spin manipulation such that GS
is a generalization of US which is a generalization of RS, as would be implied
by the successive restriction of the wave function to be an eigenfunction of
$S_z$ and $S^2$. Such a structure is possible through a change of spin basis
to the Pauli spin basis \cite{DBWY_PHD,Li18_169},
\begin{equation}
\label{eq:pauli_sep}
X^{2C} = \sum_{I=0}^3 X^I \otimes \sigma_I,
\end{equation}
where 
\begin{subequations}
\begin{align}
2X^0 &= X^{\alpha\alpha} + X^{\beta\beta}, \\
2X^1 &= X^{\alpha\beta} + X^{\beta\alpha}, \\
-2iX^2 &= X^{\alpha\beta} - X^{\beta\alpha}, \\
2X^3 &= X^{\alpha\alpha} - X^{\beta\beta}, 
\end{align}
\end{subequations}
and $\{\sigma_I\}$ is the set of Pauli matrices with $\sigma_0$ representing
the 2x2 identity matrix. $X^0$ is referred to as the  scalar component of the
spinor operator $X^{2C}$, while $X^1$, $X^2$ and $X^3$ are referred to
as its spin components. In this basis, specialization from GS to US to RS
simply implies the removal of certain spin components of density and Fock
matrices.  Thus, in ChronusQ, spin-dependent operators are stored internally as
collections of scalar and spin components according to
\begin{equation}
\label{eq:spin_generalization}
F = \begin{cases}
F^0 \otimes \sigma_0 + 
F^1 \otimes \sigma_1 + 
F^2 \otimes \sigma_2 + 
F^3 \otimes \sigma_3 & \mathrm{GS} \\
F^0 \otimes \sigma_0 + 
F^3 \otimes \sigma_3 & \mathrm{US} \\ 
F^0 \otimes \sigma_0 & \mathrm{RS}
\end{cases}
\end{equation}

The use of this scheme for spin separation is primarily a mechanism to
enable a readable and expressive code structure void of fringe logic to handle
the manipulation of the different spin cases. Internally, the eigenvalue
problems solved for the  Roothaan-Hall equations are exactly those given 
in \cref{eq:rh_1c,eq:rh_2c}. 
In the case of the molecular orbital matrix
and canonical orbital eigenvalues, no such simplification can be made. However,
due to the fact that ChronusQ is primarily a single-reference quantum chemistry 
code, the majority of the functionality may be implemented through the use of
the density matrix as opposed to the orbitals directly. Thus the vast majority
of the code base benefits from this spin separation scheme. 
The manipulation of the cases for the eigenvalue
problem is facilitated through templated matrix utilities which efficiently
convert between the two formats, which yields only minimal effect on the 
overall structure of the code base. 
In practice, we have found the following points to be the primary motivation
for the chosen approach for the treatment of spin in ChronusQ:
\begin{enumerate}

  \item Generalization of a method to support two-component spinors is
  systematic. If a method may be expressed in the Pauli spin basis, generalization
  is typically as simple as performing some set of mathematical operations on the scalar component
  followed by another set of (possibly different) operations on the spin components (e.g. \cref{eq:2c_fock})

  \item The off-diagonal matrices $X^{\alpha\beta}$ and $X^{\beta\alpha}$
  are generally non-Hermitian, while $X^{2}$ and $X^{3}$ are Hermitian
  as long as $X^{2C}$ is Hermitian. Thus when constructing moieties such
  as $K(X^{i})$ (see \cref{eq:2c_fock}), the symmetry of $X^{i}$ may be utilized; reducing the
  number of floating point operations in direct contraction schemes.

  \item The manipulation of general (possibly spin dependent) perturbations as
  they relate to the Hamiltonian and the evaluation of molecular properties is
  straight forward.  All perturbations which are spin independent (such as
  electric multipoles, electron-nuclear interaction, etc.) only appear in the
  scalar Fock matrix, while those which are spin dependent (such as magnetic
  fields) only appear in the spin components of the Fock matrix. Further, due to
  the fact that the Pauli matrices are traceless, the expectation value of a general
  one particle spinor operator, $O^\mathrm{2C}$, with respect to the one particle spinor density matrix, $P^\mathrm{2C}$, may be 
  expressed as \cite{DBWY_PHD,Li18_169}
  \begin{equation}
    \label{eq:2c_expectation}
    \langle O^\mathrm{2C} \rangle = \mathrm{Tr} \left[O^\mathrm{2C} P^\mathrm{2C}\right] = 
    2\sum_{I=0}^3 \mathrm{Tr} \left[O^I P^I\right]
  \end{equation}
\end{enumerate}

\section*{\sffamily \Large CURRENT PROGRAM FEATURES}

\subsection*{\sffamily \large Ground State Methods}

\subsubsection*{\sffamily \normalsize Reference Wave Functions}

ChronusQ implements a general self--consistent field (SCF) module which
provides a unified treatment of RS, US, and GS spin symmetries to obtain
ground state wave functions described by either Hartree--Fock (HF) or
Kohn--Sham density functional theory (KS) approximations. Within the general
framework for the treatment of spin described in the previous section, the
HF Fock matrix may be expressed as \cite{DBWY_PHD,Li18_169}
\begin{subequations}
\label{eq:2c_fock}
\begin{align}
&F^0 = h^0 + J(P^0) - K(P^0), \\
&F^k = h^k  -K(P^k), \qquad k \in \{1,2,3\},
\end{align}
\end{subequations}
where $h$ is the core Hamiltonian, and $J$ and $K$ are the Coulomb and exchange
matrices, respectively. Thus the treatment of RHF, UHF and GHF
may be expressed simply by forming additional exchange terms for the required
spin components of the density matrix. The formation of the Fock matrix is
based on the direct SCF scheme of Alml\"{o}f, \emph{et al}. \cite{Korsell82_385}
using a integral screening scheme resembling that of the LinK method
\cite{HG98_1663}.

The formation of the KS Fock matrix in ChronusQ adopts a form analogous to
\cref{eq:spin_generalization,eq:2c_fock} for the RKS, UKS and GKS methods \cite{DBWY_PHD,Li18_169}.  For
the use of two-component spinor density, ChronusQ utilizes a set of torque free
auxiliary variables which allow the adaptation of collinear XC functionals for
use with non-collinear densities \cite{Li18_169,Li17_2591}. 
This non-collinear DFT formalism preserves the self-consistent, global
zero-torque theorem for the XC magnetic field \cite{Gyorffy01_206403} while
admitting the ability to yield non-zero local torque. In practice, it is
the local torque which allows for the accurate description of spin
dynamics and time-evolution of the spin density
\cite{Gyorffy01_206403,Gyorffy03_354,Gross07_196405,Frisch07_125119,Frisch12_2193,Scuseria13_035117}.
The XC integration and batching
scheme adopted in ChronusQ closely resembles that of Stratmann, \emph{et al}. \cite{Frisch96_213}.
The assembly of the XC potential in ChronusQ generalizes the scheme of
Burow, \emph{et al}. \cite{Sierka11_3097} for use with open-shell and spinor densities \cite{Li18_169}.
ChronusQ supports local density (LDA), generalized gradient (GGA), and hybrid
approximations for the XC functional. The formation of both the HF and KS Fock
matrix is efficiently parallelized using both inter- and intranode
communication schemes. 
Ref~\citenum{Li18_169} demonstrates the linear scaling (strong and weak) of the KS
implementation in ChronusQ.

\subsubsection*{\sffamily \normalsize Relativistic Effects and External Magnetic Fields}

Relativistic effects may be included variationally in the SCF procedure for
both HF and KS wave functions through the exact two-component (X2C) method
\xcref.  Practically, the extension of the SCF module to implement the X2C-HF
and X2C-KS methods simply amounts to the manipulation of additional spin
components for the core Hamiltonian in \cref{eq:2c_fock}.  To ensure numerical stability, relativistic
calculations are performed using Gaussian charge models \cite{Dyall97_207} for nuclear
charge distributions rather than the point charge distributions typically used
in non-relativistic theory. 
For calculations including one-electron spin orbit contributions to the
core Hamiltonian, Boettger scaling \cite{Boettger00_7809} is used to approximately account for the
higher-order two-electron terms.
ChronusQ also
supports SCF optimization in the presence of finite magnetic fields through the
use of gauge including atomic orbitals (GIAO) \cite{London37_397,Ditchfield72_5688} at the
GIAO-RHF \cite{Helgaker08_154114}, GIAO-UHF, and GIAO-GHF levels of theory \cite{Tellgren18_184112,Li19_348}.
For the formation of the GIAO Fock matrix, the magnetic
field manifests both as explicit terms added to the spin component of the Fock
matrix as well as an implicit dependence of the CGTO basis function through the
addition of a complex phase factor,
\begin{equation}
\chi^\mathrm{CGTO}(\mathbf r - \mathbf R) \mapsto \exp\left( \frac{i}{2} (\mathbf R \times \mathbf B) \cdot (\mathbf r - \mathbf R) \right)
\chi^\mathrm{CGTO}(\mathbf r - \mathbf R),
\end{equation}
where $\chi^\mathrm{CGTO}$ is a CGTO centered at $\mathbf R$ and $\mathbf B$ is the external
magnetic field.  In addition, ChronusQ includes orbital--Zeeman, spin--Zeeman, and
diamagnetic field interactions of the non-relativistic Pauli Hamiltonian in the GIAO
methods \cite{Jorgensen91_2595,Helgaker08_154114,Tellgren18_184112,Li19_348}.
GIAO Fock matrices may be formed directly in the AO basis through a generalization of the
direct SCF method of Alml\"{o}f, \emph{et al}. \cite{Korsell82_385} for complex integrals.
To the best of the
authors' knowledge, ChronusQ contains the only publicly available
implementation of two-component spinor GIAO methods \cite{Li19_348} at this time.

\subsubsection*{\sffamily \normalsize Self-Consistent Field Optimization}

ChronusQ implements both first and second order optimization schemes for
the ground state SCF wave function. First order SCF optimization utilizes the
widely adopted \cite{McDouall91_167,Neese12_73,Neese18_e13127,Psi4} scheme involving a combination
of damping and direct inversion in the iterative subspace (DIIS) extrapolation of
the Fock and density matrices to accelerate convergence. Second order optimization
may proceed by either the quadratically convergent SCF method of Bacskay \cite{Bacskay81_385},
the stability check and perturbation method \cite{Pople77_3045,Li15_154109},
or some combination of the two. Typically, the best results for the optimization 
of GSCF wave functions in ChronusQ involves first order optimization of the wave
function to a stationary saddle point followed by second order optimization to
the minimum energy solution. For a full description of the SCF control options
supported by ChronusQ, we refer the reader to the ChronusQ user manual.

\subsection*{\sffamily \large Real-Time Electronic Dynamics}

One of the hallmarks of the functionality contained in ChronusQ is the
implementation of real-time propagation for quantum systems in the presence of
external perturbations, as well as its general implementation of such
functionality with respect to the treatment of electronic spin. 
Real-time propagation is important in the study of
matter-field interaction due to the fact that it is non-perturbative relative
to the strength of the external perturbation; both linear and non-linear
effects may be captured through proper analysis \cite{Rehr07_154114,Saalfrank07_034107,Li13_064104,Li18_e1341,Yabana19_094101}. 
This class of real-time
functionality is encapsulated in the RT module of ChronusQ.  For a complete description
of the specification of options for the RT module in ChronusQ, we refer the
reader to the ChronusQ user manual. 

Fundamentally, the
RT module solves the non-linear Louiville-von Neumann equation \cite{Frensley90_745,Mukamel95_9355},
\begin{equation}
i\hbar \partial_t P(t) = [F(P(t),t), P(t)]. \label{eq:lvn}
\end{equation}
Formally, \cref{eq:lvn} may be solved (propagated) exactly in the time
domain through the Magnus expansion of the time-domain propagator \cite{Magnus54_649},
\begin{equation}
P(t) = U(t, t_0) P(t_0) U(t,t_0)^\dagger, \quad U(t,t_0) = \exp \left(\Omega(t,t_0)\right),
\end{equation}
where $\Omega(t,t_0)$ is a non-terminating series expansion which must be
truncated in practice. Many explicit and implicit time integration methods may
be derived from the Magnus expansion of the propagator \cite{Rubio04_3425,Casas16_book,Li18_e1341} of which
ChronusQ implements a functional subset. The default integration scheme in
ChronusQ is the modified midpoint unitary transformation (MMUT) method of Li,
\emph{et al}. \cite{Li05_233}. MMUT is a symplectic multi-step (leap-frog),
explicit integration scheme based on the Magnus expansion with error formally
$\mathcal{O}(\Delta t^2)$,
\begin{equation}
P(t_{k+1}) = U_\mathrm{MMUT}(t_k) P(t_{k-1}) U^\dagger_\mathrm{MMUT}(t_k), 
\qquad U_\mathrm{MMUT}(t_k) = \exp \left( -\frac{2i\Delta t}{\hbar} F(P(t_k), t_k) \right).
\end{equation}
The leap-frog character of MMUT exhibits an excellent trade off between accuracy and efficiency
over other Magnus-based expansions for the oscillatory solutions which are characteristic
of these types of simulations.
Because MMUT is a leap-frog method, a non-leap-frog step must be used to seed and restart the 
integration every so often to maintain accuracy and stave off energy drift \cite{Li05_233}. For this
purpose, the default seed and restart step in ChronusQ is the explicit second order (trapezoidal) 
Magnus (EM2) propagator given by \cite{Casas16_book}
\begin{align}
P(t_{k+1}) = U_\mathrm{EM2}(t_k) P(t_k) U^\dagger_\mathrm{EM2}(t_k), \quad & 
U_\mathrm{EM2}(t_k) = \exp\left( -\frac{i\Delta t}{\hbar} F_m(t_k,t_{k+1})\right),
\end{align}
where
\begin{align}
&F_m(t_k,t_{k+1}) = \frac{1}{2} ( F(P(t_k),t_k) + F(P_\mathrm{FE}(t_{k+1}),t_{k+1}) ), \\
&P_\mathrm{FE}(t_{k+1}) = U_\mathrm{FE}(t_k) P(t_k) U^\dagger_\mathrm{FE}(t_k), \quad 
U_\mathrm{FE}(t_k) = \exp\left( -\frac{i\Delta t}{\hbar} F(P(t_k), t_k) \right).
\end{align}
For both the MMUT and EM2 integration methods, the matrix exponential may be formed using
either exact diagonalization based methods, or approximate Taylor series and Chebyshev series 
expansions \cite{Li16_5333,Rubio04_3425,Kosloff91_59}. The latter approximate approaches yield drastic performance improvements
for large quantum systems/basis sets due to the poor scaling and parallelization of eigenvalue
decomposition.

ChronusQ admits the ability to express an arbitrary number of (one-body) 
time-dependent external perturbations, $\{V_\xi(t)\}$, of the form
\begin{equation}
F(P(t),t) = F_0(P(t)) + \sum^{N_\mathrm{pert}}_\xi V_\xi(t), \label{eq:td_pert}
\end{equation}
where $F_0$ is only implicitly time-dependent through the time dependence of the 
density matrix. Generally, a particular perturbation takes the functional form
\begin{equation}
V_\xi(t) = V_\xi \cdot f(t) (\Theta( t - t_\mathrm{on}) - \Theta( t - t_\mathrm{off})),
\end{equation}
where $\Theta$ is the Heaviside step function, $t_\mathrm{on/off}$ are reference
times to turn the perturbation on and off, $V_\xi$ is a physical, time-independent
one-body operator, and $f(t)$ is an envelope function which describes the time
dependence and amplitude of the perturbation. For $V_\xi$, ChronusQ supports both the length and
velocity gauge electric multipole operators up through the electric octupole,
and the magnetic dipole operator.  Additionally, ChronusQ supports the
following envelope functions:
\begin{equation*}
f(t) = \kappa \cdot
\begin{cases}
  1                                              & \text{Step Function} \\
  t - t_\mathrm{on}                              & \text{Linear Ramp} \\
  \cos (\omega(t - t_\mathrm{on}))               & \text{Plane Wave} \\
  \exp \left(-\alpha (t-t_\mathrm{on})^2 \right) & \text{Gaussian}
\end{cases}
\end{equation*}
where $\kappa$ is a chosen parameter to control the amplitude of the perturbation. An example field specification is given in \cref{fig:cq_input}.

Throughout the time-propagation, ChronusQ evaluates the time-dependent expectation
values of various one body operators, $\{ O_\zeta \}$, according to
\begin{equation}
    \langle O_\zeta (t) \rangle = \mathrm{Tr} \left[O_\zeta P(t)\right]. \label{eq:td_expectation}
\end{equation}
Depending on the spin characteristics of $O_\zeta$, \cref{eq:td_expectation}
may be efficiently evaluated according to \cref{eq:2c_expectation}. For
example, evaluation of a time-dependent component (i.e. the $z$ component) of
the electric dipole amounts to only evaluating the trace of the $z$-dipole
matrix with respect to the scalar density matrix, while evaluating a component
of the ($z$) component of the magnetic dipole requires evaluating the trace of
the $z$-magnetic dipole with respect to $P^3(t)$.  The 
supported choices for $O_\zeta$ are the same as those for $V_\xi$ outlined above. 
The time-series for the desired time-dependent expectation values may be
obtained from the ChronusQ checkpoint file after the calculation 
for post processing at the users' discretion. Using these time-series, users may simulate 
quantities such as absorption spectra \cite{Li18_e1341,Li17_5283}, 
electric  and magnetic circular dichroism spectra \cite{Li16_234102,Li19_accepted},
charge transfer \cite{Li14_24457,Li14_244705,Li16_7255}, 
and second harmonic generation \cite{Li13_064104,Barron04_book}.

\subsection*{\sffamily \large Frequency Domain Methods}

In addition to time-domain methods, ChronusQ implements a suite of frequency
domain methods to solve time-dependent problems in the RESP module. For a full
description of the RESP module and its input specification, we refer the reader
to the ChronusQ user manual.  As opposed to explicit, non-perturbative solution
of \cref{eq:lvn} in the time domain, these methods solve \cref{eq:lvn} in the
frequency domain under a series of approximations, namely that the applied
perturbation is weak in field strength and harmonic in the time domain. Under
these approximations, solution of \cref{eq:lvn} (or more specifically \cref{eq:td_expectation})
instead becomes a large linear
algebra problem to obtain the quantum propagator in the frequency domain.
Frequency domain methods fall into one of two classes: residue methods and response
methods. In this section, we review the salient aspects of the implementation
of these methods in ChronusQ.


\subsubsection*{\sffamily \normalsize Residue Methods}

The quantum propagator diverges in the frequency domain at frequencies
which correspond to transition energies of the quantum system \cite{Yeager84_33}.
Residue methods aim to determine the poles and residues of the propagator
at these frequencies, thus obtaining information pertaining to the excited states
of the quantum system such as transition energies and densities. 
For the single determinant reference wave functions implemented in ChronusQ,
residue methods generally take the form of a large generalized eigenvalue problem
of a matrix pencil $(H,S)$,
\begin{equation}
HT = ST\Omega, \label{eq:res}
\end{equation}
where the form of $H$ and $S$ are dependent on the representation of the reference
wave function (i.e. in the case of HF/KS response, $H$ represents the orbital Hessian
and $S$ is an indefinite metric \cite{Yeager84_33,Casida95_Book,HeadGordon05_4009}).  
In general, the columns of the eigenvectors $T$ represent the transition density
pertaining to the transition energies of the corresponding diagonal element of
$\Omega$. 
ChronusQ supports residue calculations using two different approximations
of the quantum propagator, namely 
particle-hole (ph-RPA) \cite{Yeager84_33,Casida95_Book,HeadGordon05_4009} and 
particle-particle (pp-RPA) \cite{Yang13_224105,Yang13_18A522,Yang13_174110}
random phase approximations.
Both HF or KS reference wave functions may be used for either of
these propagator approximations. Determining propagator poles and residues under the ph-RPA yields information
relating to neutral (particle number conserving) excitations of the quantum system
in question, while those corresponding to the pp-RPA yield information pertaining
to two electron addition (double electron affinity) and removal (double ionization).
Additionally, ChronusQ supports inclusion of both relativistic
effects (the X2C-ph-RPA \cite{Li16_3711} and X2C-pp-RPA \cite{Li16_5379} methods) and
magnetic fields (the GIAO-ph-RPA method \cite{Li19_3162,Tellgren19_accepted}) in its residue calculations.
The GIAO-ph-RPA method using GIAO-RHF and GIAO-GHF reference wave functions have enabled the calculation of linear
response approximations for the magnetic circular dichroism spectra of closed shell systems \cite{Li19_3162}.
To the best of the authors' knowledge, ChronusQ contains the only publicly 
available implementation of the GIAO-ph-RPA method at this time.
We note for clarity that the GIAO-ph-RPA method, as implemented in ChronusQ, treats
the linear response of molecular systems to external perturbations in the presence of explicit finite, 
static magnetic fields. We are careful to make this distinction in this work to avoid 
confusion with work on perturbative GIAO response methods such as those developed by Kj{\ae}rgaard, \emph{et al} 
\cite{Michl07_11278,Coriani09_1997}.

Generally, the dimension of $H$ is too large to directly diagonalize using
methods such as those provided in (Sca)LAPACK. For these large problems,
ChronusQ implements the iterative generalized preconditioned locally harmonic
residual (GPLHR) method \cite{Krylov15_273,Xue16_A500} to perform the diagonalization. Rather than
obtaining all possible eigenvalues of $(H,S)$, the GPLHR method performs a partial
diagonalization for a select number of desired eigenvalues.  Unlike the more
typically utilized Davidson (Lanczos) method \cite{Davidson75_87,Scott86_817,Morgan92_287}, GPLHR is able to directly (and
robustly) solve for eigenpairs in the spectral interior of $(H,S)$, thus making
it an ideal solver for high-energy spectroscopies such as X-Ray absorption
\cite{Li18_2034}. Like the Davidson method, GPLHR only requires the specification
of action (product) of $(H,S)$ onto a trial set of vectors to perform the 
diagonalization.  ChronusQ implements direct, density based contraction schemes 
to perform the matrix products for both the ph-RPA \cite{Helgaker00_8908} and pp-RPA \cite{Yang13_174110}
methods. Because transition densities exhibit the same spin factorization 
as the density matrix (\cref{eq:pauli_sep}) \cite{Li16_3711}, the direct contractions closely
resemble those of \cref{eq:2c_fock} in the AO basis, only with a non-Hermitian density operator.
As such, the generalization of the residue calculations to use RS, US and GS
reference wave functions closely resembles the structure of the Fock matrix formation.

\subsubsection*{\sffamily \normalsize Linear Response Methods}

Evaluation of the residues may be avoided entirely if the moiety of interest
is \cref{eq:td_expectation} in the frequency domain. Within the linear
response (weak perturbation) approximation, frequency dependent properties may
be evaluated directly through the solution of a shifted system of linear equations
given by \cite{Agren06_143001,Norman10_5096,Oddershede01_JCP}
\begin{subequations}
\begin{align}
  \langle O_\zeta \rangle = \langle\langle O_\zeta; V_\xi \rangle\rangle^\gamma_\omega \approx 
  \mathbf{O}_\zeta^\dagger \mathbf{X}_\xi(\omega),  \label{eq:gf} \\
  ( H - (\omega + i\gamma) S)\mathbf{X}_\xi(\omega) =  \mathbf{V}_\xi. \label{eq:lr}
\end{align}
\end{subequations}
Here, $O_\zeta$ and $V_\xi$ are defined as they were in \cref{eq:td_expectation,eq:td_pert}, respectively,
and $(H,S)$ is the same pencil as \cref{eq:res}. $\gamma$ is a complex damping parameter which
ensures convergence of the propagator at all frequencies (including poles) \cite{Norman10_5096}. The 
boldface $\mathbf{O}_\zeta$ and $\mathbf{V}_\xi$ are property gradients defined by the selection
of the propagator approximation and operators $O_\zeta$ and $V_\xi$, respectively \cite{Yeager84_33}. ChronusQ
currently supports linear response calculations using the ph-RPA only.
Much like the residue methods, the dimension of $H$ is often too large to factorize
or invert directly using (Sca)LAPACK. As such, ChronusQ provides an implementation
of the generalized minimum residual (GMRES) method \cite{Walker88_152} to iteratively
solve large linear systems of the form of \cref{eq:lr}. 

In addition to providing solvers which solve \cref{eq:lr} in the full dimension
of $H$, ChronusQ implements a series of model order reduction (MOR) techniques \cite{Li17_4950,Yang19_accepted} to improve
the efficiency of evaluating \cref{eq:lr} for multiple frequencies in a spectral region
of interest. The MOR technique constructs a rational Krylov subspace which interpolates
\cref{eq:gf} in the frequency domain. The subspace is constructed as the linear span
of solutions to \cref{eq:lr} at multiple, carefully chosen interpolation frequencies.
Using Galerkin projection under the constructed subspace, the full dimensional problem 
is projected onto a reduced space problem which is rapidly  evaluatable. The MOR
method implemented in ChronusQ has been demonstrated to provide excellent approximations
to the X-Ray absorption spectrum for large molecular systems with quadratic computational scaling. ChronusQ was
the first software to provide such MOR functionality and remains the sole publicly 
available implementation at this time to the best of the authors' knowledge.

\section*{\sffamily \Large CURRENT DEVELOPMENTS}

As is the case with many research codes, there are a number of development streams
which have not yet made it into the primary release of ChronusQ. In this section,
we examine some of the most recent development in the ChronusQ ecosystem which
will be folded into forthcoming releases.

\subsection*{\sffamily \large Kohn--Sham Density Functional Theory in Magnetic Fields}

Magnetic density functional theory (BDFT) strikes a balance between cost and accuracy for simulating molecules in
magnetic fields \cite{Harris94_3089}. Currently, the released version of ChronusQ only supports GIAO basis functions
for the treatment of finite magnetic fields within the GIAO-HF level of theory. GIAO-RKS reference
wave functions \cite{Helgaker14_034101,Helgaker17_4089,Stopkowicz19_97} are currently supported in the development 
version of ChronusQ, and will be present in the forthcoming release.

\subsection*{\sffamily \large Quantum Embedding Methods}

In recent years, quantum embedding capabilities have been developed in ChronusQ
\cite{Li19_174114}. Currently, users may perform an embedded SCF calculation by
specifying both an embedding potential and embedding magnetic field on a real
space grid using the XSF file format \cite{Kokalj03_155}. Such moieties have regularly
been obtained using a combination of the Abinit \cite{Zwanziger09_2582} software and the
software presented in previous work by Huang, \emph{et al.} on embedding \cite{Carter11_154110}.  
CGTO basis representations of the embedding
perturbations are produced by numerical integration of the interpolated real
space representations using a similar scheme as the one used for the XC
integration for KS methods in ChronusQ (for details, see \cite{Li19_174114}).
Embedded SCF calculations are supported for RSCF, USCF, and GSCF optimizations
for both HF and KS levels of theory. Embedding X2C-HF/KS wave functions is
currently under development.

\subsection*{\sffamily \large Two-Component Post-SCF Methods}

The development version of ChronusQ includes working implementations of
both ground- and excited-state post-SCF methods. Among the ground-state
post-SCF methods, standard second-order M{\o}ller-Plesset perturbation theory (MP2)
\cite{Frisch88_503}, approximate second-order coupled-cluster
(CC2) \cite{Jazrgensen95_409}, and coupled-cluster with single and double
excitations (CCSD) \cite{Bartlett82_1910} have been implemented. For
post-SCF excited states, ChronusQ includes an implementation of the time-dependent (TD)
equation-of-motion coupled-cluster
(EOM-CC) \cite{Mukherjee79_325,Emrich81_379,Stanton93_7029} formalism of
Nascimento and DePrince \cite{Nascimento16_5834,Nascimento17_2951}, in both 
the TD-EOM-CC2 and TD-EOM-CCSD flavors.

Unlike other reported time-dependent coupled-cluster
implementations \cite{Schonhammer78_6606,Kvaal12_194109,Huber11_054113,Sato18_051101,Pedersen19_144106},
the TD-EOM-CC formalism in ChronusQ is based on the time evolution of a
coupled-cluster dipole moment function from which one can obtain a
spectral line shape function, $I_\xi(\omega)$, defined as
\begin{subequations}
\begin{align}
  I_{\xi}(\omega) =  \int\limits_{-\infty}^{\infty} dt\,e^{-i\omega t} \langle \Phi_0|(\hat{1}+\hat{\Lambda})\bar{\mu}_\xi e^{i\bar{H}_Nt} \bar{\mu}_\xi|\Phi_0 \rangle , ~ \xi \in \{x,y,z\}. \label{eq:td-eom-cc2}
\end{align}
\end{subequations}
Here, $\bar{\mu}_\xi$ represents the $\xi$-component of the similarity-transformed electric dipole
operator 
\begin{equation}
\bar{\mu}_\xi = e^{-\hat{T}} \hat{\mu}_\xi e^{\hat{T}}
\end{equation}
and $\bar{H}_N$ represents the normal-ordered similarity-transformed Hamiltonian
\begin{equation}
\bar{H}_N = e^{-\hat{T}}\hat{H}e^{\hat{T}} - E_{\rm CC}.
\end{equation}
The symbols $E_{\rm CC}$, $\hat{T}$, and $\hat{\Lambda}$ represent the energy,
excitation operator, and de-excitation operator associated with the ground-state CC
wavefunction, respectively. The real part of the line shape function defines oscillator strengths
\begin{equation}
\label{EQN:OSCILLATOR_STRENGTH}
f(\omega) = \frac{2}{3}\omega \sum_\xi {\Re}\{I_\xi(\omega)\}.
\end{equation}
which may be evaluated over an
arbitrarily broad energy window \cite{Nascimento16_5834,Nascimento17_2951}.

A notable feature of the ground- and excited-state implementations
discussed here is that they rely on the same template metaprogramming techniques
employed throughout the rest of ChronusQ (see Appendix) and thus are
compatible with both real and complex RHF, UHF, and GHF (including X2C-HF)
reference wavefunctions.  Hence, like many
other reported implementations of time-independent relativistic EOM-CC
theory \cite{Klein08_194106,Epifanovsky15_064102,Cheng18_044108,Badala18_21051,Pathak14_042510,Pathak16_94,Pathak16_074110,Akinaga17_827,Avijit18_174113,Wang14_5567,Zheyan12_174102,Yang12_236,Ayush19_074102,Nguyen19_1423,Liu18_144108},
the X2C-based TD-EOM-CC2 and TD-EOM-CCSD algorithms in ChronusQ
incorporate spin-orbit and scalar relativistic effects from first
principles. To the best of the authors' knowledge, this implementation is the only one that
incorporates spin-orbit coupling within a time-dependent EOM-CC formalism at this time.
Active effort is also being undertaken toward the development of standard
(frequency domain) EOM-CC methods \cite{Stanton93_7029} within ChronusQ. 

\section*{\sffamily \Large CONCLUSIONS}

ChronusQ is a standalone, high-performance quantum chemistry software 
package specializing in time-dependent and relativistic electronic 
structure methods. Special emphasis is placed on the general treatment of 
electronic spin throughout the code base, allowing for the implementation
of electronic structure methods with various spin symmetries while requiring only 
minor code modifications. Further, the modern C++ code design adopted by 
ChronusQ allows for the expression of complex quantum chemistry algorithms
in a readable, generic, and elegant manner. Thus, as a development code, 
ChronusQ has proven to be a near optimal platform for the rapid development
of novel electronic structure methods. In addition, with extensive exploitation
of high performance numerical linear algebra software, and use of both
shared and message passing parallelism schemes, ChronusQ is a performant,
production quality quantum chemistry code which is suitable for both workstations and 
supercomputers alike.

In this article, we reviewed much of the salient aspects of the functionality
implemented in the ChronusQ program as well as a number of the current
developments which will accompany its upcoming release. The core of the
functionality in ChronusQ revolves around the solution of time-dependent 
electronic structure problems in either the time or frequency domain.
ChronusQ supports the inclusion of relativistic effects in the electronic
wave function through the X2C method throughout the code. This
support in enabled in part by the general manner with which ChronusQ
treats electronic spin, as outlined in the main text. Further, ChronusQ
supports the inclusion of finite, uniform magnetic fields as external
perturbations in all HF based electronic structure methods. Implementations
of methods such as GIAO-GHF and the GIAO-ph-RPA in ChronusQ are, at the time this article has been written and to the best
of the authors' knowledge, the sole implementations of these methods in
publicly available, open source quantum chemistry codes. Further,
ChronusQ provides implementations of cutting edge numerical linear
algebra algorithms such as the GPLHR method, and MOR approximations
for problems arising in linear response theory. To the best of the authors'
knowledge, ChronusQ provides the sole implementation of MOR functionality
for linear response problems available to the public.

ChronusQ is a \emph{free} (as in \emph{software freedom}) quantum chemistry
software package which is open to the public. Contributions from the
quantum chemistry community as a whole are not only allowed, but encouraged.
ChronusQ may be obtained online through our website \cqweb. This website
also contains the ChronusQ Wiki and user manual which serve as the most
up-to-date references for the functionality contained in ChronusQ.

\section*{\sffamily \Large ACKNOWLEDGMENTS}

The development of Chronus Quantum ChronusQ computational
software is supported by the National Science Foundation
(OAC-1663636 to X.L. and A.E.D.). The method development of
magneto-optical spectroscopy is supported by National Science
Foundation (CHE-1856210 to X.L.), and the development of
relativistic electronic structure theory is supported by the
Office of Science, US Department of Energy (DE-SC0006863 to X.L.).
E.V. was partially supported by the U.S. National Science Foundation 
(award 1550456).

\section*{\sffamily \Large AUTHOR CONTRIBUTIONS}

The overall structure and modern C++ design of ChronusQ is due to David B.
Williams-Young (DBWY). Parallelism in ChronusQ is also due to DBWY. The name
``Chronus Quantum" is due to Joshua J. Goings (JJG). The development of memory
management utilities in ChronusQ is due to DBWY,  Torin F. Stetina (TFS), and
Andrew Wildman (AW).  The development of the ground state SCF module is due to
DBWY, Patrick Lestrange (PL), TFS, and Xiaosong Li (XL). The development of DFT
related methods is due to Alessio Petrone (AP), DBWY and TFS.  ChronusQ
primarily utilizes the Libint library to evaluate molecular integrals over
contracted Gaussian basis functions. The Libint library is due to Edward F.
Valeev (EFV), and its integration into ChronusQ is due to DBWY and EFV.
ChronusQ also contains an in house integral code which performs molecular
integral evaluations for those integrals which are not provided by Libint, such
as those required by the GIAO and X2C methods.  The in house integral code is
due to Shichao Sun (SS) and XL.  The development of GIAO capabilities in
ChronusQ are due to SS and XL.  The development of the X2C method in ChronusQ
is due to DBWY and Joseph Kasper.

The RT module is due to DBWY, AP, AW, SS, JJG, and Feizhi
Ding.  The MMUT algorithm implemented in the RT module is due to XL.  The RESP
module and the development of the MOR capabilities in ChronusQ are due to DBWY.
The extension of DFT and GIAO capabilities to the RESP module are due to AP and
SS, respectively. Validation of the RESP module is in large part due to TFS
and AW.

The development of embedding methods in ChronusQ is due to Chad E. Hoyer. 
The development of post-SCF methods such as Coupled Cluster and 
M{\o}ller-Plesset perturbation theory is due to DBWY, Daniel R. Nascimento,
Lauren Koulias, and A. Eugene DePrince III. The tensor contractions required
by these post-SCF methods are efficiently performed by the TiledArray library, 
which is due to EFV.


\subsection*{\sffamily \Large APPENDIX: MODERN C++ SOFTWARE DESIGN}

The adoption of modern C++ in ChronusQ has allowed for a level
of abstraction throughout the code base which has enabled 
the development of expressive, reusable and efficient code.
ChronusQ was originally written in C++11, but more recently we have adopted a number of
features in the C++14 standard (such as generic lambda expressions, variable templates, etc.)
to decrease verbosity and improve expressiveness throughout the code. Current development
streams of ChronusQ require features from C++17 (such as polymorphic memory resources,
template deduction, etc.); however, these features will not be merged into the release 
stream until the standard is more widely supported in vendor C++ compilers. In this
appendix we review a number of the more salient aspects on the modern C++ design
of ChronusQ.

\subsection*{\sffamily \large Template Metaprogramming}

The use of template metaprogramming in ChronusQ is primarily motivated by the
fact that we wish to develop code which supports both real and complex wave
functions with maximal code reuse and minimal duplication of type-generic 
implementation details. To this end, the use of templates allows for
the specialization of a certain number type-specific implementation details
while allowing the developer to express their algorithms in a type-generic
manner.

%
%

In electronic structure theory, the majority of the type-specific
implementation details are encapsulated in the use of linear algebra software.
Such operations include matrix multiplication, diagonalization, etc.
Implementations of the BLAS and LAPACK standards are typically expressed in
either the C or FORTRAN programming languages such that their calling APIs are
type-specific, i.e. the multiplication of two real (double precision) matrices
(DGEMM) is performed using a different function than the one that performs the
multiplication of two complex (double precision) matrices (ZGEMM). However, the
generic structure of these APIs (i.e. the general information one needs to pass
to perform the operation such as data pointers, dimensions, etc.) are invariant
between the function calls. To this end, ChronusQ implements templated wrappers
around a functional subset of the BLAS and LAPACK standards to provide uniform
calling API which allows for their integration into type-generic algorithmic
structures. In addition, ChronusQ also provides type-generic templated wrappers
for the MPI API to a similar effect. We note for completeness that similar
functionality is currently being developed and standardized in the BLAS++,
LAPACK++, and SLATE software libraries \cite{SLATE}. There are currently
development efforts to integrate such software into ChronusQ. 

\subsection*{\sffamily \large Type Independent API and Dynamic Polymorphism}

\begin{figure}
\lstset{language=C++,
                basicstyle=\tiny\ttfamily,
                keywordstyle=\color{blue}\ttfamily,
                stringstyle=\color{red}\ttfamily,
                commentstyle=\color{magenta}\ttfamily,
                morecomment=[l][\color{magenta}]{\#},
                frame=single
}
\begin{center}
\begin{tabular}{c}
\begin{lstlisting}[linewidth=6.2cm]
// Type independent API specification
class SingleSlaterBase {
  // Type Independent API
  void SCF(); // calls formFock, etc.
  ...
  // Type Dependent API
  virtual void formFock(...)        = 0;
  virtual void contractOrbHess(...) = 0;
  ...
};

// Type-generic SingleSlater implementation
template <typename WfnT, typename IntT>
class SingleSlater : public SingleSlaterBase { 
  // Internal storage, etc.
  ...
  // Type specific implementations
  void formFock(...)        override { ... };
  void contractOrbHess(...) override { ... };
  ...
};

\end{lstlisting} 
%
\end{tabular}
\end{center}
\caption{Code snippet which demonstrates the separation of type independent algorithmic specification 
and type generic implementation for the \texttt{SingleSlater} class in ChronusQ. \texttt{WfnT} and \texttt{IntT} represent the 
data types describing the electronic wave function and molecular integrals, respectively.
}
\label{fig:singleslater}
\end{figure}

Due to the fact that ChronusQ supports many combinations of methods throughout
the code base (i.e. non-relativistic, time-dependent, X2C, GIAO, etc.), the
majority of the core functionality must be templated over both a type which
describes the wave function and a type which describes the basis functions to
maintain a type-generic development platform. However, because C++ is strongly
typed, providing a type-generic API for type-agnostic operations such as
performing the self consistent field (SCF) optimization would require the calling program to implement
a combinatorial number of paths to perform these operations for every allowable
combination of template parameters.  To solve this problem in ChronusQ, we have
adopted an object oriented code structure which separates type agnostic
interfaces from type dependent implementations through the use of dynamic
polymorphism.

To demonstrate this state of affairs, we examine the code snippet in
\cref{fig:singleslater} which describes the implementation of the \texttt{SingleSlater}
class in ChronusQ.  \texttt{SingleSlater} implements the minimum necessary
functionality (i.e.  formation and storage of the Fock matrix, contractions
with the orbital Hessian for response calculations, etc.) for the description of the electronic wave
function as a single Slater determinant.  
In this example, \texttt{SingleSlaterBase} provides a type erased API which is
accessible from the calling program to perform type agnostic operations.
Additionally, \texttt{SingleSlaterBase} provides the required API
specification for functionality in which the underlying implementation must be
type dependent, i.e. Fock matrix formation and storage, etc. These functions are left as pure virtual to ensure that any
derived class must provide their implementations. Using the derived implementations
of the type dependent operations, one may implement operations such as 
the SCF optimization in a type agnostic manner. 



\bibliography{Journal_Short_Name,rel-refs,linalg_ref,qc_software,Li_Group_References,Petrone,dbwy_refs,Embedding,GIAO_refs,post-scf}
\end{document}